\documentclass[aps,nofootinbib,preprintnumbers,twocolumn,superscriptaddress,prd]{revtex4-1}

\usepackage{amsmath}
\usepackage{amssymb}
\usepackage{slashed}
\usepackage{epsfig}
\usepackage{bbm,bm}
\usepackage{hyperref}
\usepackage{color}
\usepackage{comment}
\usepackage{amsfonts}
\usepackage{dsfont}

\usepackage{multirow}

\makeatletter
\@addtoreset{equation}{section}

\makeatother

\def\bea{\begin{eqnarray}} 
\def\eea{\end{eqnarray}}
\def\be{\begin{equation}} 
\def\ee{\end{equation}} 
\def\ba{\begin{array}}
\def\ea{\end{array}} 
\def\nn{\nonumber}

\def\nn{\nonumber}

\let\oldtitle\title
\renewcommand{\title}[1]{\oldtitle{\color{blue}{#1}}}

\let\oldeqref\eqref
\let\oldcite\cite
\renewcommand{\eqref}[1]{{\color{blue}\oldeqref{#1}}}
\renewcommand{\cite}[1]{{\color{blue}\oldcite{#1}}}

\makeatletter
\let\reftagform@=\tagform@
\def\tagform@#1{\maketag@@@{\ignorespaces\textcolor{blue}{(\ignorespaces #1 \unskip\@@italiccorr \ignorespaces)\ignorespaces}}}
\makeatother

\makeatletter
\renewcommand{\p@subsection}{}
\renewcommand{\p@subsubsection}{}
\makeatother

\usepackage{mathpazo}

\begin{document}

\title{
On critical models with $N\leq 4$ scalars in $d=4-\epsilon$
}

\author{A.\ Codello}
\email{a.codello@gmail.com}
\affiliation{Instituto de F\'isica, Faculdad de Ingenier\'ia, Universidad de la Rep\'ublica, 11000 Montevideo, Uruguay}

\author{M.\ Safari}
\email{mahsafa@gmail.com}
\affiliation{Romanian Institute of Science and Technology, Str.~Virgil Fulicea  3, 400022 Cluj-Napoca, Rom\^ania}

\author{G.\ P.\ Vacca}
\email{vacca@bo.infn.it}
\affiliation{INFN - Sezione di Bologna, via Irnerio 46, 40126 Bologna, Italy}

\author{O.\ Zanusso}
\email{omar.zanusso@unipi.it}
\affiliation{Universit\`a di Pisa and INFN - Sezione di Pisa, Largo Bruno Pontecorvo 3, I-56127 Pisa, Italy}

\begin{abstract}
We adopt a combination of analytical and numerical methods to study the renormalization group flow
of the most general field theory with quartic interaction in $d=4-\epsilon$ with $N=3$ and $N=4$ scalars.
For $N=3$, we find that it admits only three nondecomposable critical points:
the Wilson-Fisher with $O(3)$ symmetry, the cubic with $H_3=(\mathbb{Z}_2)^3\rtimes S_3$ symmetry,
and the biconical with $O(2)\times \mathbb{Z}_2$.
For $N=4$, our analysis reveals the existence of new nontrivial solutions with discrete symmetries
and with up to three distinct field anomalous dimensions.
\end{abstract}

\maketitle

\section{Synopsis and summary of our results}

Critical models are theories in which the correlation length diverges
thanks to the precise tuning of some external parameters.
Assuming the thermodynamical limit, the criticality condition
ensures that no scale has any effect on the theory
because it is infinitesimally small
when compared to the diverging correlation length. This results in
a model which has no built-in scale and thus is allowed to exhibit fluctuations
of arbitrary size with a fractal-like behavior.
Importantly, critical models can be studied as the scale invariant
fixed points of renormalization group (RG) beta functions \cite{Wilson:1973jj}.
In this regard, quantum and statistical field theories
are the most commonly used frameworks and, especially, scalar theories
with local interactions in $d=4-\epsilon$ dimensions
have been the subject of several investigations since almost half a
century~\cite{Brezin:1973jt,Zia:1974nv,Michel:1983in,TLTB}.

The leading and next-to-leading orders of the perturbative renormalization group flow
of the model
\be \label{eq:lagrangian}
\mathcal{L} = {\textstyle{\frac{1}{2}}} \, \partial\varphi  \cdot \partial\varphi +  v(\varphi)\,,
\ee
with the
most general quartic potential
\begin{equation}\label{eq:general-potential}
  \begin{split}
   v(\varphi)= \frac{1}{4!} \lambda_{ijkl} \varphi_i\varphi_j\varphi_k\varphi_l
  \end{split}
\end{equation} 
in $d=4-\epsilon$ dimensions is given by
the following beta functional (all dimensionful quantities are rescaled to be dimensionless and an additional rescaling removes a factor $(4\pi)^2$)
\begin{equation}\label{eq:nlo-flow}
  \begin{split}
    \beta_v &= -d v + \frac{d-2}{2} \varphi_i v_i + \gamma_{ij} \varphi_i v_j + \frac{1}{2} v_{ij} v_{ij}
    - \frac{1}{2} v_{ij} v_{ikl} v_{jkl} \\
    \gamma_{ij} &= \frac{1}{12} v_{ikln}v_{jkln} -\frac{1}{16}v_{iklm} v_{jknp} v_{lmnp}\,,
  \end{split}
\end{equation}
in which $\varphi_i$ with $i=1,\dots,N$ are scalar fields, subscript indices on $v(\varphi)$ stand for field derivatives,
$\gamma_{ij}$ is the anomalous dimension matrix,
and repeated indices are summed over.
Scale invariant solutions $v^*(\varphi)$ of $\beta_v=0$ lead to perturbative fixed points of the renormalization group for which $\lambda^*_{ijkl} \sim {\cal O}(\epsilon)$,
and correspond to critical models with nontrivial behavior for $d<4$.
Eigenvalues of the $\gamma$ matrix are fields' anomalous dimensions;
upon diagonalization they are related
to the anomalous dimensions $\eta_i$ of the two point functions,
$2\gamma_{ij}\sim\delta_{ij}\eta_i$ (no summation over the index $i$).

The flow \eqref{eq:nlo-flow} is known to the next-to-next-to-leading order \cite{Osborn:2017ucf}.
For finding perturbative fixed points, it is sufficient to use the leading order of \eqref{eq:nlo-flow};
however, we displayed the next-to-leading term for two main reasons:
on the one hand, it is universal among massless RG schemes;
on the other hand, it is necessary for determining whether any found solution
actually corresponds to a ${\cal O}(\epsilon)$ degenerate set of physically distinct fixed points.
The latter situation is known to occur sometimes \cite{Michel:1983in,TLTB}.
Furthermore, the compact functional form of the RG flow presented here is very convenient since it is able to describe
easily the scaling properties of composite operators and some operator product expansion (OPE)
coefficients~\cite{ODwyer:2007brp,Codello:2017hhh}. 
Simple and powerful relations to the conformal field theory (CFT) description
have been derived at leading order both for single ~\cite{Rychkov:2015naa,Nii:2016lpa,Codello:2017qek} 
and multi-scalar field systems~\cite{Codello:2018nbe,Codello:2019vtg}.
Notice that, strictly speaking, scale invariance does not imply conformal invariance \cite{Polchinski:1987dy}.
Here, we concentrate on solutions of \eqref{eq:nlo-flow} and assume that, if the stronger condition
of conformal invariance is met, then some of the CFT data can be deduced by RG analysis.

The RG flow can be written as the gradient of a function $A$ with a metric in couplings' space
which is trivial in the parametrization \eqref{eq:general-potential} up to the next-to-leading order \cite{Wallace:1974dy}. Using the couplings $\lambda_{ijkl}$ and their beta functions $\beta_{ijkl}$,
the function $A$ is obtained by integrating $\beta_{ijkl} \delta \lambda_{ijkl} = \delta A$, resulting at the leading order in
\be \label{eq:a-function}
A=-\frac{\epsilon}{2} \lambda_{ijkl}\lambda_{ijkl}+\lambda_{ijmn}\lambda_{mnkl}\lambda_{klij}\,.
\ee
At a fixed point $\lambda^*_{ijkl}$ and at the leading order (LO), $A$ can be related to the anomalous dimensions 
\be
A^*\stackrel{\mathrm{LO}}{=}-\frac{\epsilon}{6} \lambda^*_{ijkl}\lambda^*_{ijkl}\stackrel{\mathrm{LO}}{=}\, -\epsilon \sum_i \eta_i \,,
\label{Avalue}
\ee 
which can be shown to have a bound $A\ge -\frac{N}{48} \epsilon^3$ that can be saturated in certain cases~\cite{Rychkov:2018vya}.
Another quantity of interest is the coefficient of the energy-momentum tensor two point function $C_T$, which can be normalized relatively to the single free scalar theory case $C_{T,free}$.
At the leading order the ratio can be written~\cite{Osborn:2017ucf} as
\be
\frac{C_T}{C_{T,free}}=N-\frac{5}{36}  \lambda^*_{ijkl}\lambda^*_{ijkl}\stackrel{\mathrm{LO}}{=}N-\frac{5}{6}\sum_i \eta_i\,.
\ee

The form of the potential $v(\varphi)$ in \eqref{eq:general-potential} is not constrained by any symmetry group acting on the fields' multiplet $\varphi_i$.
The maximal symmetry that the potential can have is $O(N)$,
which would act on the fields by rotating them
$\varphi_i \to \varphi_i'=(R\cdot \varphi)_i = R_i{}^j\varphi_{j} $ with $R\in O(N)$.
Even in absence of maximal symmetry, the action of $O(N)$ on the fields induces an action on the potential itself
$v(\varphi) \to v'(\varphi) \equiv v(R\cdot \varphi)$ and thus on the couplings $\lambda_{ijkl}$,
which can generally be decomposed using the irreducible representations (irreps)
of $O(N)$. This method allows us to rewrite the very complicate and rather redundant fixed point equations of the couplings $\lambda_{ijkl}$
in terms of simpler equations that depend on new coupling with irreducible $O(N)$ transformations.
The simpler equations are then more suitable to be studied either analytically, or numerically, or with a combination of both methods.

For $N=1$ there is only one critical model with $O(1)\simeq\mathbb{Z}_2$ symmetry, the well-known $\phi^4$ model,
which captures the physics of the universality class of the Ising model \cite{Wilson:1973jj}.
Similarly, for $N=2$ there is only one critical model with $O(2)$ symmetry,
which captures the physics of the universality class of the Heisenberg model \cite{Brezin:1972se}. In this latter case and for the rest of this paper,
we have restricted our attention to nondecomposable models,
that is, to fixed points which cannot be written as simple sums
of models with lower values of $N$.
The statement for $N=1$ is very easy to show since it involves only one coupling,
while the case for $N=2$ can be proven by brute force specializing the RG beta
functions \eqref{eq:nlo-flow}, or, more sophisticatedly,
using the representation theory of  the group $O(2)$ as done in Ref.~\cite{Osborn:2017ucf}.

There are three known solutions for $N=3$.
The two better-known ones have anomalous dimension matrix
$\gamma_{ij}$ proportional to the identiy; they are
the Wilson-Fisher model with maximal $O(3)$ symmetry \cite{Brezin:1973jt},
and the cubic model with $H_3$ symmetry \cite{Aharony:1973zz,Aharony:1973,Carmona:1999rm}.
The lesser-known solution has two different anomalous dimensions
as diagonal entries of $\gamma_{ij}$;
it is the biconical model with $O(2) \times \mathbb{Z}_2$ symmetry \cite{Kosterlitz:1976zza,Calabrese:2002bm}.
Interestingly, all these $N=3$ solutions
admit generalizations to arbitrary values of $N$~\cite{Rychkov:2018vya}.
To the best of our knowledge, there is no proof that the three aforementioned solutions for $N=3$
are the complete set of ${\cal O}(\epsilon)$ solutions to $\beta_v=0$.

In Sect.~\ref{section:N=3} we adopt the method of the irreps of $O(3)$
and combine it with analytical and numerical methods to solve $\beta_v=0$.
To summarize our first important result here,
we do not find any further solution for $N=3$, besides the ones mentioned
in the previous paragraph.
Therefore, our analysis strongly suggests
that the only solutions for $N=3$ are the Wilson-Fisher $O(3)$,
the cubic, and the biconical fixed points.

Much less is known for models with $N\geq 4$.
As previously mentioned,
there are some families of critical models which generalize the ones seen for $N=3$
as a function of $N$.
These include the $O(N)$ Wilson-Fisher \cite{LeGuillou:1977rjt}, the (hyper)cubic with
$H_N = (\mathbb{Z}_2)^N\rtimes S_N$ symmetry,
the (hyper)tetrahedral with $\mathbb{Z}_2 \times S_{N+1}$,
but also several others.
All these solutions are characterized by one independent field anomalous dimension,
with the exception of the biconical solution with $O(N_1)\times O(N_2)$ symmetry and $N=N_1+N_2$,
which is characterized by two distinct field anomalous dimensions. 
For a rather complete accounting of these solutions, we refer to \cite{Rychkov:2018vya}.
The numerical methods that we have applied for $N=3$ can be
straightforwardly generalized to $N\geq 4$, however the equations quickly
increase in size and complexity by orders of magnitude,
so we do not attempt an empirical classifications of critical
models for $N=4$ and beyond.

There are, however, results that we believe are interesting on the basis
of our experience with the irreps method and of some observations.
The observation is rather empirical: roughly, the \emph{more} symmetry a critical model has,
the \emph{less} independent anomalous dimensions it has. Groups such as $O(N)$ and $H_N$
fully constrain the matrix $\gamma_{ij}$ to be proportional to the identity, because they have at
most one quadratic invariant.
For both $N=2$ and $N=3$, it seems that the critical models can have at most
$N-1$ independent $\eta$s (the biconical model being the first one with two independent $\eta$s for $N=3$).
This leads us to the conjecture that for a general $N$-components model
there can be at most $N-1$ independent anomalous dimensions. 
We set off numerically testing and confirming this hypothesis for $N=4$ in Sect.~\ref{section:N=4}.
Therefore, the situation appears to be different from what happens for scalar theories in $d=6-\epsilon$ dimensions,
for which, in a previous work, we have shown that $N$-components models exist with $N$ independent $\eta$s up to $N=3$~\cite{Codello:2019isr}.  

We report three new solutions for $N=4$ with discrete symmetries in Sect.~\ref{section:N=4}.
One of them has two independent anomalous dimensions,
the other two have three independent anomalous dimensions.

\section{Tensor decomposition}\label{tensor_dec}

A simple and transparent way to decompose the couplings in irreps of $O(N)$, which works for any number of flavors $N$,
is to split the symmetric tensor $\lambda_{ijkl}$ as follows
\begin{equation}
\begin{split}
  \lambda_{ijkl} = \kappa\delta_{(ij}\delta_{kl)}+\rho_{(ij}\delta_{kl)}+\sigma_{ijkl}\,, \quad \rho_{ll}=0\,, \quad \sigma_{ijll}=0 \,,
\end{split}
\end{equation}
where $\rho_{ij}$ and $\sigma_{ijkl}$ are symmetric traceless tensors and indices enclosed curved brackets are symmetrized.
Taking the trace of this relation one finds, for general $N$, that
\begin{equation}
\begin{split}
& \kappa = \frac{3}{N(N + 2)}\lambda_{aabb}\,,
\quad
\rho_{ij} = \frac{6}{N + 4}\Bigl(\lambda_{ijaa}-\frac{1}{N}\delta_{ij}\lambda_{aabb}\Bigr) \,, \\
& \sigma_{ijkl} = \lambda_{ijkl} - \kappa\delta_{(ij}\delta_{kl)}-\rho_{(ij}\delta_{kl)} \,.
\label{tensordec}
\end{split}
\end{equation}
The main advantage of this decomposition is that it immediately suggests how redundant couplings can be eliminated.
We first notice that if $v^*(\varphi)$ is a solution of $\beta_v=0$, then any rotation of the solution,
such as $v^{*\prime}(\varphi)=v^*(R\cdot \varphi)$ for $R\in O(N)$, is also a solution.
The solution might be invariant under the rotation, such as the case of the $O(N)$ model,
or transform nontrivially into another one, which belongs to the same moduli space of solutions
with completely equivalent critical properties.
Therefore, one can always apply a $O(N)$ rotation without changing the physics.
We argue that it is best to use such freedom to bring
the matrix $ \rho_{ij} $ to diagonal form,
hence reducing the number of couplings in the model and greatly simplifying the search in theory space. 

If we further impose some additional constraints, for example that some elements on the diagonal of $ \rho_{ij} $ are the same,
we would then have extra rotational freedom left that is not broken by the form of $ \rho_{ij} $,
which can thus be used to set to zero some components of $ \sigma_{ijkl} $. 
In the next two sections, we discuss the applications of these methods to the case $ N=3 $ and $N = 4$.

\section{Irreducible results for $N=3$}\label{section:N=3}

By restricting to a specific value of $ N $,
one can make an explicit connection between the components of the irreps and the couplings of the model.
For the special case of $ N=3 $, there are $15$ independent couplings among $\lambda_{ijkl}$,
which we denote as $ \lambda_I $, $ I=1, \cdots, 15$.
We thus rewrite \eqref{eq:general-potential} as
\bea\label{eq:potential-N=3}
 v &=& \frac{1}{4!} \Bigl(
\lambda _1 \varphi _1^4 +4 \lambda _2 \varphi _1^3 \varphi _2  +4 \lambda _3 \varphi _1^3 \varphi _3  +6 \lambda _4 \varphi _1^2 \varphi _2^2  
 \nn\\
&&
  +6 \lambda _7 \varphi _1^2 \varphi _3^2 + 12 \lambda _5 \varphi_1^2 \varphi _2 \varphi _3 
  +4 \lambda _6  \varphi _1 \varphi _2^3+4 \lambda _{12} \varphi _1 \varphi _3^3 
 \nn\\
&& + 12 \lambda _{10} \varphi _1 \varphi _2 \varphi _3^2 +12 \lambda _8 \varphi _1 \varphi _2^2 \varphi _3 
 +\lambda _9 \varphi _2^4 +\lambda_{15} \varphi _3^4
 \nn\\
&& + 4 \lambda _{14} \varphi _2 \varphi _3^3+6 \lambda _{13} \varphi _2^2 \varphi _3^2+4 \lambda _{11} \varphi _2^3 \varphi _3
\Bigr)
\eea
The $ \lambda_I $ carry a reducible representation of $ SO(3) $, that is decomposed as
\be 
15 = 1 \oplus 5 \oplus 9
\label{irreps3}
\ee
corresponding to the $ \kappa $, $ \rho_{ij} $ and $ \sigma_{ijkl} $ representations described earlier.
To find the explicit linear combinations of $ \lambda_I $ that make up these irreps,
one simply diagonalizes simultaneously $ J^2=J \cdot   J $ (the quadratic Casimir) and $ J_3 $ operators in their representation
on $ \lambda_I $ space, where $ J_i $ are the $so(3)$ algebra generators satisfying $ [J_i, J_j] = i \epsilon_{ijk} J_k $. 

Doing this, we can list the components $g_I$ of the irreps of the right-hand side of \eqref{irreps3} as linear combinations of $\lambda_I$.
The components of the $9$ dimensional irrep are $ g_I $, $ I=1, \cdots, 9 $ and are found to be
\begin{equation}
\begin{split}
g_1 &= \sqrt{2} \left(3 \left(\lambda _1+2 \lambda _4-8 \lambda _7+\lambda _9-8 \lambda _{13}\right)+8 \lambda _{15}\right) \\
g_2 &= 30 \left(\lambda _5+\lambda _{11}\right)-40 \lambda _{14} \\
g_3 &= 40 \lambda _{12}-30 \left(\lambda _3+\lambda _8\right) \\
g_4 &= -10 \left(\lambda _1-6 \lambda _7-\lambda _9+6 \lambda _{13}\right) \\
g_5 &= -20 \left(\lambda _2+\lambda _6-6 \lambda _{10}\right) \\
g_6 &= 70\left(\lambda _{11}-3 \lambda _5\right) \\
g_7 &=  70 \left(\lambda _3-3 \lambda _8\right) \\
g_8 &= 35 \left(\lambda _1-6 \lambda _4+\lambda _9\right) \\
g_9 &=  140 \left(\lambda _2-\lambda _6\right)\,.
\end{split}
\end{equation}
The $5$ dimensional irrep in \eqref{irreps3}, denoted by $ g_I $ for $ I = 10, \cdots, 14 $ is
\begin{equation}
\begin{split}
g_{10} &= \sqrt{2}\left(\lambda _1+2 \lambda _4-\lambda _7+\lambda _9-\lambda _{13}-2 \lambda _{15}\right) \\
g_{11} &=  6 \left(\lambda _5+\lambda _{11}+\lambda _{14}\right) \\
g_{12} &= -6 \left(\lambda _3+\lambda _8+\lambda_{12}\right) \\
g_{13} &=  -3 \left(\lambda _1+\lambda _7-\lambda _9-\lambda _{13}\right) \\
g_{14} &=  -6 \left(\lambda _2+\lambda _6+\lambda _{10}\right)\,.
\end{split}
\end{equation}
Finally, the last linear combination gives the singlet representation 
\begin{equation}
\begin{split}
g_{15} = \lambda _1+2 \lambda _4+2 \lambda _7+\lambda _9+2 \lambda _{13}+\lambda _{15} \,.
\end{split}
\end{equation}

As discussed earlier, one can remove some redundant couplings by a suitable rotation and diagonalize the tensor $\rho_{ij}$
of Sect.~\ref{tensor_dec}.
For our choice of $ g_I $ basis,
the diagonalization of $ \rho_{ij} $ is simply equivalent to setting $g_{11}=g_{12}=g_{14}=0$. 
Moreover $4$-dim subspaces $\{g_1,g_{10},g_{15}, g_i\}$ for $i=6,7,8,9$ are invariant under the RG flow.

A numerical analysis of the reduced system of beta functions shows
that there is no new fixed point beyond what is already known in the literature,
namely, the Wilson-Fisher $ O(3) $, the cubic and the biconical fixed points.
The $3$ by $3$ diagonal matrix $ \rho $ vanishes completely for the $O(3)$ and cubic fixed points,
while it has two equal diagonal nonzero elements for the biconical.
For the $ O(3) $ model the only nonzero coupling is the singlet $ g_{15} = 15\epsilon/11  $, because it has maximal symmetry.
For the cubic fixed point the nonzero couplings are $ g_{1} = -14\sqrt{2}\epsilon/9  $, $ g_{8} = -70 \epsilon/9 $
and $g_{15} = 4\epsilon/3 $.
Finally, for the biconical fixed point, analytic expressions for the coupling values are available,
but too long to present here, instead we give the numerical values
$ g_{1} = -2.6617 \epsilon$, $ g_{10} = 0.05529 \epsilon$ and $g_{15} = 1.3371 \epsilon$.

Now for some details our our search procedure:
We first use the global rotational freedom to set $g_{11}=g_{12}=g_{14}=0$ on all beta functions;
a fixed point is thus defined as a solution of the overcomplete set of equations ($15$ equations
in $12$ variables). Since the system is still beyond analytic treatment,
we first scan analytically several subsystems by setting some further  $g_i$ to zero,
which allows us to find analytic expressions for the three solutions.
Then, we construct an algorithm that repeatedly searches for numerical zeroes
of the full system from random initial conditions. These initial conditions are
chosen first within the Rychkov-Stergiou (RS) bound \cite{Rychkov:2018vya},
which takes the form of an ellipsoid in couplings' space,
and then within a (hyper)rectangular box containing the RS bound itself.

\section{New nontrivial $N=4$ critical theories}\label{section:N=4}

Let us now consider the case $N=4$.
We want to devote the investigation of this section to the search of previously unknown critical theories.
The details of the decomposition in irreps for $N=4$ are discussed in Appendix~\ref{irrepsNge4}.
To summarize the most important facts,
one can define the basis of couplings $g_I$ from the eigenstates of the (only independent) Casimir $c$
and of the two operators $h_1$ and $h_2$ of the Cartan subalgebra.
As for $N=3$, the beta functions of $g_I$ greatly simplify the analysis of the fixed points.
Likewise the previous section, we find useful to choose a specific basis for which the $\rho_{ij}$ matrix of the tensor decomposition~\eqref{tensordec} is diagonal,
which can always be done by applying a suitable $O(4)$ transformations.

In the following, we focus our attention on the search of solutions that do not possess the so-called trace property \cite{Brezin:1973jt,Pelissetto:2000ek},
i.e.\ with  nonzero $\rho_{ij}$.\footnote{This property has been intensively investigated in the past because it guarantees
that criticality can be tuned by a single parameter.}
Taking into account the corresponding set of six conditions,
one can consider also different sets of nonzero $g_i$ couplings,
which are characterized by the eigenvalues of the operators $c$, $h_1$ and $h_2$.
Then one can see which of them are the solutions of the apparently overconstrained system of equations
given by the complete set of beta functions, as explained in the previous section.

Since we do not rely on the assumption of any symmetry,
this approach can be considered a kind of tamed almost brute force procedure, which nevertheless is very effective.
Combining numerical and analytical algorithms to solve the system of polynomial equations, we can easily find several solutions.
Clearly, we are able to find all the well-known families of fixed points and their symmetry properties specialized to $N=4$,
that we already recalled in the Introduction.
These solutions have been largely discussed in the literature, also recently~\cite{Osborn:2017ucf, Rychkov:2018vya},
and we do not address them here. 
We instead concentrate on three new nontrivial fixed point solutions.
One of them is characterized by two distinct field anomalous dimensions, and the other two by three different ones.
All these solutions have discrete global symmetries.

The three new critical theories have LO potentials
\begin{widetext}
\begin{eqnarray}
v_1/\epsilon &=& a_1
   \left(\varphi _1^2+\varphi _4^2\right){}^2+a_2 \left(\varphi _2^2+\varphi _3^2\right){}^2+a_3 \left(\varphi
   _1^2+\varphi _4^2\right) \left(\varphi _2^2+\varphi _3^2\right)
+ a_4 \left(\varphi _3 \varphi_1^3-3 \varphi _1 \left(\varphi _1 \varphi _2+\varphi _3 \varphi _4\right) \varphi _4+\varphi _2 \varphi _4^3\right) \,, \nonumber \\
v_2/\epsilon &=& b_1 \left(\varphi _1^4+\varphi_3^4\right)+b_2 \varphi _2^4+b_3 \varphi _4^4+b_4 \varphi _1^2 \varphi _3^2+b_5 \varphi _2^2 \varphi _4^2
+b_6 \left(\varphi _1^2+\varphi
   _3^2\right) \varphi _2^2+b_7 \left(\varphi_1^2+\varphi _3^2\right) \varphi _4^2+b_8 \varphi _1 \varphi _2 \varphi _3\varphi_4 \,,\nonumber \\ 
v_3/\epsilon &=& c_1 \varphi _1^4+c_2 \varphi _2^4+c_3 \varphi _1^2 \varphi _2^2+c_4 \left(\varphi _3^2+\varphi _4^2\right) \varphi _1^2+c_5 \varphi _2^2 \left(\varphi _3^2+\varphi _4^2\right)
+c_6 \left(\varphi _3^2+\varphi_4^2\right){}^2 +c_7 \varphi _1 \varphi_4 \left(\varphi _4^2-3 \varphi _3^2\right) \, ,
\label{potentials}
\end{eqnarray}
with coefficients
\begin{eqnarray*}
&& a_1=0.00919041\,,\quad a_2=0.00970702\,,\quad a_3=0.0232858\,,\quad a_4=0.00383258\,, \quad
 b_1=0.00836122\,, \quad b_2=0.00977968
\\&&
 b_3=0.0120982\,, \quad b_4=0.0298269\,, \quad b_5=0.018415\,,\quad
 b_6=0.0235331\,,\quad b_7=0.014846\,, \quad b_8=0.0229149\,,
\\&& 
 c_1=0.009894\,,\quad c_2=0.0117361\,, \quad c_3=0.0261495\,, \quad c_4=0.0192239\,, \quad
     c_5=0.0106247\,,\quad c_6=0.0112214\,,
\\&&
 c_7=-0.00450868 \nonumber\,.
\end{eqnarray*}
\begin{center}
	\begin{table} [bt]
		\begin{tabular}{| l | c | c | l | r | r |} 
			\hline
			 & $\eta_i/\epsilon^2$  & A$ _*/\epsilon^3$  & Symmetry & $I_2$ & $I_4$  \\ \hline \hline
			$v_1$ & $(0.0210892, 0.0210892, 0.0205446, 0.0205446)$  &  -0.0832676 & $\mathbb{Z}_3 \times \mathbb{Z}_2^2$  
  ~\hfill see also\textsuperscript{\ref{footnote:v1}}
 & 2 & 4 \\ \cline{1-6}
			$v_2$ & $(0.0212805, 0.0208156, 0.0205473, 0.0205473)$  &  -0.0831906 & $\mathbb{Z}_2^4$ 
  ~\hfill see also\textsuperscript{\ref{footnote:v2}}
 & 3 &  8 \\ \cline{1-6}
			$v_3$ & $(0.0212688, 0.020709, 0.020709, 0.0204991)$      & -0.0831859  & $S_3 \times \mathbb{Z}_2^2 $                & 3 &  7  \\ \hline
		\end{tabular}
		\caption{Anomalous dimensions $\eta_i$, values of $A$ at LO, apparent symmetry and number of quadratic and quartic invariants of the novel $N=4$ critical theories.} 
\label{table1}
	\end{table}
\end{center}
\end{widetext}
We have given rounded numerical values of the critical couplings, instead of analytic expressions,
because they are roots of polynomials with cumbersome expressions.
All three solutions are bounded from below,
which can be seen by first arranging them as $v \sim \sum_{ab}{\cal O}_a Q_{ab} {\cal O}_b$,
for some monomials ${\cal O}_a$ quadratic in the fields, and by then testing the positivity of the matrices $Q_{ab}$.

In Table~\ref{table1}, we present for each new critical theory the numerical values of the anomalous dimensions $\eta_i$,
which combine to $A$ at LO using \eqref{eq:a-function},
their apparent symmetries, and the number of quadratic and quartic invariants, respectively denoted $I_2$ and $I_4$.
Notice that the number of quadratic invariants $I_2$ equals the number of independent anomalous dimensions.

In the chosen basis, the tensorial decomposition ~\eqref{tensordec} gives the following matrix $\rho_{ij}/\epsilon$ 
for the three fixed points
\begin{eqnarray}
v_1 &:& 0.00619928 \, \, {\rm diag} (-1,1,1,-1) \,,\nonumber\\
v_2 &:& {\rm diag} (-0.006082,-0.006082, 0.01128, 0.0008868) \,, \nonumber\\
v_3 &:& {\rm diag} (-0.001903, -0.001903, 0.01112, -0.007316)\,, \nonumber 
\end{eqnarray}
 %
%
implying that the first matrix has symmetry $O(2)^2$, while the second and third ones $O(2)$.
These are not symmetries of the critical models, however,
because the traceless tensor $\sigma_{ijkl}$, which has several nonzero elements for all three critical theories,
breaks them down to the discrete ones shown in Table~\ref{table1}.

Invariants of various order can be also constructed from $\kappa$, $\rho_{ij}$ and $\sigma_{ijkl}$, which provide
some degree of information on the properties of the solutions.
A useful feature of the solutions is obtained by looking at the spectrum of the quadratic and quartic invariants,
from which we can count the singlet scaling operators with respect the symmetry of the specific critical theory.
The invariants do not need to be scaling operators and one can combine operators with different anomalous dimensions and obtain invariants with higher symmetry, up to $O(4)$, e.g.\ $\sum_i \varphi_i^2$ and $(\sum_i \varphi_i^2)^2$.
For the quadratic operators, we start from the eigenvalue equation for the scaling operators
$S_2=S_{ij}\varphi_i \varphi_j$, given by $\gamma_2 S_{ij}= v_{ijab} S_{ab}$ as described in~\cite{Codello:2019vtg}.
Moreover for the quartic operators, we simply investigate the spectrum of the stability matrix of the beta functions at criticality.

In the following we briefly summarize our findings for the case $N=4$:
\begin{itemize}
\item $v_1$: the symmetry $\mathbb{Z}_3 \times \mathbb{Z}_2^2$ reflects the invariance under the simultaneous exchange ($\varphi_1\leftrightarrow \varphi_4$, $\varphi_2\leftrightarrow \varphi_3$), under the sign changes $(\varphi_2, \varphi_4) \to (-\varphi_2, -\varphi_4)$ and $\mathbb{Z}_3$ transformations in the $(\varphi_1,\varphi_4)$ sector.\footnote{%
Addendum: after this paper has been published, it has been shown in~\cite{Osborn:2020cnf} that this fixed point has a larger $O(2)$ symmetry which,
given the potential as in Eq.~\eqref{potentials}, is realized as $\phi_1 \to \phi_1 \cos{\theta}+\phi_4 \sin {\theta}\,, \,\phi_4 \to -\phi_1\sin {\theta}+\phi_4 \cos {\theta} \,, \,
\phi_2 \to \phi_2 \cos {3 \theta}+ \phi_3 \sin {3 \theta} \,,\,\phi_3 \to -\phi_2 \sin {3 \theta} +\phi_3 \cos {3 \theta}$. \label{footnote:v1}}
There are two singlet quadratic operators with scaling dimension $2-0.488829\epsilon$ and $2-0.115889\epsilon$,
both invariant under the symmetry, implying that $I_2=2$.
Counting the quartic singlet scaling operators one finds $I_4=4$.
There are $18$ IR-attractive, $12$ IR-repulsive and $5 $ marginal deformations at LO in the quartic sector.\footnote{%
Here and for the following two solutions, ``marginal at LO'' means that the deformations do not have a ${\cal O}(\epsilon)$ contribution.
They do have a ${\cal O}(\epsilon^2)$ contribution, instead, which comes from the next-to-leading term of \eqref{eq:nlo-flow}.
In other words, they do not parametrize a conformal manifold \cite{Behan:2017mwi}.}
\item $v_2$: it is characterized by the symmetry $\mathbb{Z}_2^4$,
because of the invariance under the exchange $\varphi_1\leftrightarrow \varphi_3$, 
and under the simultaneous sign change $\varphi_i \to -\varphi_i$ for the three independent pairs $(12)$, $(13)$ and $(24)$.\footnote{%
Addendum: because of the nontrivial homomorphism (induced by the reflection) a direct product becomes semi-direct and it has been observed in~\cite{Osborn:2020cnf} that the symmetry should be written as $D_4 \times \mathbb{Z}_2$ where $D_4=\mathbb{Z}_2^2 \rtimes \mathbb{Z}_2$. \label{footnote:v2}}
There are $6$ quadratic scaling operators with different eigenvalues (one dimensional eigenspaces),
but only three of them respect the symmetry, implying that $I_2=3$. The quadratic singlets have critical exponents
$2-0.481864\epsilon$, $2-0.226530 \epsilon$, and $2-0.136650\epsilon$.
There are $8$ quartic singlet scaling operators, i.e.\ $I_4=8$.
We find $16$ IR-attractive, $13$ IR-repulsive and $6$ marginal deformations at LO in the quartic sector.
\item $v_3$: it appears to have the symmetry $S_3 \times \mathbb{Z}_2^2$, since the potential is $S_3$ invariant in the  $(\varphi_3,\varphi_4)$ sector,
but also does not change under the transformations
$(\varphi_1, \varphi_4)  \to  (-\varphi_1, -\varphi_4)$ and $\varphi_2  \to  -\varphi_2$.
There are $4$ quadratic scaling operators with different eigenvalues (one dimensional eigenspaces) and three of them respect the symmetry with eigenvalues
$2-0.481257\epsilon$, $2-0.257752\epsilon$, $2-0.139196\epsilon$, implying $I_2=3$. 
The quartic singlet scaling operators that respect the symmetry are $7$, i.e.\ $I_4=7$.
There are $15$ IR-attractive, $14$ IR-repulsive and $6$ marginal deformations at LO in the quartic sector.
\end{itemize}
Obviously, one can straightforwardly push the analysis of these critical theories to higher orders in the perturbative expansion.
We also find, but not report, a fixed point solution related to a theory with three different 
real and positive field anomalous dimensions, but with three complex couplings,
implying that the underlying theory is nonunitary.


\section{Conclusions and future prospects}

In this short note, we have reviewed how one can conveniently carry on the search of general critical scalar theories
without assuming any symmetry, but instead letting it emerge at criticality.
The method takes advantage of the decomposition of the couplings, that define the theory space,
in irreducible representations of $O(N)$.
In a previous work we have applied the same method to critical theories with $N=3$ scalars in $d=6-\epsilon$ dimensions~\cite{Codello:2019isr}, 
providing the whole set of possible scaling solutions.
In this work, instead, we have concentrated our attention to critical scalar theories in $d=4-\epsilon$ dimensions.

For $N=3$, our study of all possible solutions,
which is based on a combination of analytical and numerical techniques, gives strong evidence 
that no new critical theories beyond the three ones that were already known do exist
($O(N)$ Wilson-Fisher, cubic and biconical models).  
Even though nothing new is found for $N=3$, this is an important finding
because it constrains the number of possible critical models with three scalar fields.

For $N=4$, we are able to find, without much numerical effort,
three previously unknown critical theories bounded from below and
characterized by nontrivial discrete symmetries.
We regard these solutions as interesting because their anomalous dimension matrix $\gamma_{ij}$ is not 
proportional to the identity.
As a consequence, one solution has two different field anomalous dimensions,
and the other two have three different ones.
It would be interesting, but maybe challenging, to investigate more these new critical theories with CFT bootstrap methods \cite{ElShowk:2012ht,Poland:2018epd}, as currently done, for example, for the cubic model \cite{Kousvos:2018rhl,Kousvos:2019hgc}.
We cannot exclude that further nontrivial scaling solutions exist, because the system of equations is rather complex, 
even after the application of the $O(4)$ reduction and the decomposition in irreducible representations.

As a final comment for future applications, especially with the interest of reaching similar results for $N \geq 5$,
we stress that the method can be easily generalized to higher values of $N$
and thus provides an important foundation for future searches.
Some notion in this direction is given in Appendix~\ref{irrepsNge4}.

The arising zoology of critical scalar models in theory space, together with their emergent symmetries,
is an interesting topic in its own right,
but it could also be useful to discover previously unknown condensed matter models
with nontrivial second order phase transitions.
It could also have some repercussion in the constructions of the Higgs sector of 
new particle physics theories.
An application of the latter idea could be relevant
for the search of new fundamental UV complete QFT within the asymptotic safety scenario (without gravity) 
along the lines discussed in Ref.~\cite{Litim:2014uca}.


\paragraph*{Acknowledgments.}
We are grateful to A.~Stergiou for correspondence related to the topic of the paper,
to S.~Rychkov for useful comments on the first version of the draft,
and to A.~Aharony for pointing out important references.


\appendix

\section{Decomposition in irreps for $N\ge 4$}\label{irrepsNge4}

We can label $\lambda_I$ the $\binom{N+3}{4}$ different couplings of the potential $v$, generalizing
the parametrization \eqref{eq:potential-N=3} to arbitrary $N$,
and study their transformation properties induced by the mixing of the monomials under $O(N)$ transformations of the fields.
The picture becomes simpler moving to a coupling basis $g_i$ suggested by the decomposition of the action of $O(N)$ in irreps,
or, even more simply, of its subgroup $SO(N)$ by leaving aside some discrete reflexion.
For simplicity, we first and foremost concentrate on the case $N=4$,
since it is the focus of our partial analysis of Sect.~\ref{section:N=4},
but also outline briefly the $N=5$ case, because it incorporates the blueprint for the generalization to arbitrary $N$.

A convenient starting point to study the irreps for $N=4$ is to choose a new field basis $\tilde\varphi_i$,
for which the Lagrangian \eqref{eq:lagrangian} takes the form
\be \label{eq:lagrangian-new-basis}
\mathcal{L} = {\textstyle{\frac{1}{2}}}\,\partial\tilde\varphi  \cdot M  \cdot \partial\tilde\varphi + \tilde v(\tilde\varphi)\,,
\ee
where 
\be 
M = \left(\ba{c|c}0&1_2\\ \hline 1_2&0\ea\right)
\ee
and both fields and potential are distinguished from those in the canonical basis of \eqref{eq:general-potential} by a tilde.
One can use the transformation matrix
\be 
X = \frac{1}{\sqrt{2}}\left(\ba{cc}1_2&-i1_2\\1_2&i1_2\ea\right) 
\ee
to move back to the canonical basis $\varphi_i$.
Here, $1_2$ is the $2$-dimensional identity matrix.
In such basis the $ SO(4) $ transformations are defined through $ O^TMO = M $.
Also, the solution to the equation $ t^TM+Mt=0 $, defining an element $t$ of the algebra, takes a simple form.

Take $e_{ij}$ to be the $4$ by $4$ matrix with ``$1$'' on the $(i,j)$ entry and ``$0$'' everywhere else.
The $ SO(4) $ algebra generators can be written as
{\setlength\arraycolsep{2pt}
\be 
\ba{lll}
h_1 &=& e_{11}-e_{33}   \\
h_2 &=& e_{22}-e_{44}    
\ea \quad
\ba{lll}
x_1 &=& e_{12}-e_{43}  \\
x_2 &=& e_{21}-e_{34}
\ea \quad
\ba{lll}
y &=& e_{14}-e_{23}  \\
z &=& e_{32}-e_{41} \,.
\ea
\ee}%
In particular, elements of the Cartan subalgebra $ h_1, h_2 $ take a diagonal form, thanks to the choice of noncanonical basis. Defining $h_{\pm} = h_1 \pm h_2 $ the nonzero commutation relations are
\bea  \label{x12}
\hspace{-1.5cm}
&&    [x_1, x_2] = h_-, \;\;
[h_-, x_1] = 2x_1, \;\;
[h_-, x_2] = -2x_2, \nonumber \\
\label{yz}
&& [y, z] = -h_+, \;\;
[h_+, y] = 2y, \;\;
[h_+, z] = -2z\,.
\eea 
One can work instead with a basis for the algebra that makes the connection $SO(4)=SO(3)\times SO(3)$ manifest:
$ [J_i, J_j] = i \epsilon_{ijk} J_k $ and $ [K_i, K_j] = i \epsilon_{ijk} K_k $. It is easily seen, from \eqref{x12}, that one has to make the identification
\be 
J_+ = \frac{x_1}{\sqrt{2}}, \qquad
J_- = \frac{x_2}{\sqrt{2}}, \qquad
J_3 = \frac{h_-}{2}, 
\ee
and similarly, from \eqref{yz},
\be 
K_+ = \frac{iy}{\sqrt{2}}, \qquad
K_- = \frac{iz}{\sqrt{2}}, \qquad
K_3 = \frac{h_+}{2}\,.
\ee
The sum of $J^2$ and $K^2$ above is equal to the Casimir operator of $SO(4)$:
\be
c = {\textstyle{\frac{1}{2}}}(h_1^2+h_2^2+\{x_1,x_2\}-\{y,z\}) = J^2 + K^2\,.
\ee 
In terms of the 
eigenvalues of the highest weight vectors under $ (h_1, h_2) $, which we denote by $(a_1,a_2)$,
and the eigenvalues of $ J^2$ and $K^2 $, denoted by $j_1(j_1+1)$ and $j_2(j_2+1)$ respectively,
the value of the Casimir $c$ is 
\be 
c = {\textstyle{\frac{1}{2}}}[a_1(a_1+2)+a_2^2] = j_1(j_1+1)+j_2(j_2+1)\,,
\ee
where $(j_1,j_2)=(\frac{a_1+a_2}{2},\frac{a_1-a_2}{2})$. 
One finds that the space of the $35$ couplings of the general $N=4$ scalar model is reduced according to the irreps 
labeled by $(j_1,j_2) = (0,0), (1,1), (2,2)$  or $(a_1,a_2)=(0,0), (2,0),(4,0)$,
with the corresponding Casimirs being $c = 0, 4, 12$. The states are labeled according to their eigenvalues under $c, h_1, h_2$.
The dimensions of the irreps are given by $(2j_1+1)(2j_2+1)=(1+a_1)^2-a_2^2$ and therefore one has the decomposition
\be 
35 = 1 \oplus 9 \oplus 25\,.
\label{irreps4}
\ee
In relation to the tensorial decomposition discussed in Sect.~\ref{tensor_dec},
one can make the identification
\be 
\kappa \leftrightarrow r_1 \quad, \quad
\rho_{ij} \leftrightarrow r_{9} \quad, \quad 
\sigma_{ijkl} \leftrightarrow r_{25}\,,
\ee
where we denoted by $r_i$ the multiplet of the $i$-dimensional representation.

One of the advantages of using the basis \eqref{eq:lagrangian-new-basis}
is that the above procedure can be extended more easily to higher $ N $.
For $ SO(5) $, we modify the previous definition by choosing
\be 
M = \left(\ba{c|c|c}0&1_2&0\\ \hline 1_2&0&0\\ \hline 0&0&1\ea\right) 
\ee
and simply add to the algebra generators of $ SO(4) $ four new elements:
{\setlength\arraycolsep{2pt}
\be 
\ba{lll}
u_1 &=& e_{15}-e_{53}   \\
u_2 &=& e_{25}-e_{54}    
\ea \quad
\ba{lll}
v_1 &=& e_{35}-e_{51}  \\
v_2 &=& e_{45}-e_{52}\,,
\ea
\ee}%
where here $e_{ij}$ is a 5 by 5 matrix, while the Cartan subalgebra remains two dimensional, consisting of $ h_1, h_2 $. The additional commutation relations would be
\be
[u_i, v_i] = -h_i, \quad
[h_i, u_i] = u_i, \quad 
[h_i, v_i] = - v_i\,.
\ee
A quadratic Casimir can be written as
\be
c ={\textstyle{\frac{1}{2}}}( h_1^2+h_2^2+\{x_1,x_2\}-\{y,z\}-\{u_1, v_1\}-\{u_2, v_2\})\,. \nonumber
\ee
Having discussed the two cases $ N=4 $ and $N=5$, the generalization to higher $ N $ (both even and odd) is straightforward.


\end{document}